# CCWSIM: An Efficient and Fast Wavelet-Based CCSIM for Categorical Characterization of Large-Scale Geological Domains

**Mojtaba Bavandsavadkoohi**
centre Eau Terre Environnement
Institut National de la Recherche Scientifique
Québec, Canada.
Faculty of Mining, Petroleum and Geophysics
Shahrood University of Technology
Shahrood, Iran.

**Erwan Gloaguen**
centre Eau Terre Environnement
Institut National de la Recherche Scientifique
Québec, Canada.

**Behzad Tokhmechi**
Faculty of Mining, Petroleum and Geophysics
Shahrood University of Technology
Shahrood, Iran.

**Alireza Arab-Amiri**
Faculty of Mining, Petroleum and Geophysics
Shahrood University of Technology
Shahrood, Iran.

**Bernard Giroux**
centre Eau Terre Environnement
Institut National de la Recherche Scientifique
Québec, Canada.

March 30, 2024

## ABSTRACT

Over the last couple of decades, there has been a surge in various approaches to multiple-point statistics simulation, commonly referred to as MPS. These methods have aimed to improve several critical aspects of realism in the results, including spatial continuity, conditioning, stochasticity, and computational efficiency. Nevertheless, achieving a simultaneous enhancement of these crucial factors has presented challenges to researchers. In the approach that we propose, CCSIM is combined with the Discrete Wavelet Transform (DWT) to address some of these concerns. The primary step in the method involves the computation of the DWT for both the Training Image (TI) and a region shared with previously simulated grids at a specific level of wavelet decomposition. Then, the degree of similarity between the wavelet approximation coefficients is measured using a Cross-Correlation Function (CCF). These approximation coefficients offer a compressed representation of the pattern while capturing its primary variations and essential characteristics, thereby expediting the search for the best-matched pattern. Once the best-matched pattern in the wavelet approximation coefficients is identified, the original pattern can be perfectly reconstructed by integrating the DWT detail coefficients through an Inverse-DWT transformation. Experiments conducted across diverse categorical TIs demonstrate simulations comparable to multi-scale CCSIM (MS-CCSIM), accompanied by an enhancement in facies connectivity and pattern reproduction. The source code implementations are available at https://github.com/MBS1984/CCWSIM.

***Keywords***: MPS simulation, Wavelet transform, Texture synthesis, Computational efficiency, Cross-Correlation, Computer graphics

## 1. Introduction

Accurate characterization and modeling of geological heterogeneity, along with associated uncertainties, are essential in quantitative resource evaluation. This importance is particularly evident when utilizing geostatistical models for estimating extractable resources in reservoirs or assessing potential hazards related to groundwater pollution. This significance is demonstrated in industries such as oil and gas, hydrogeology, and mining [1].

The limitations of variogram-based geostatistics in characterizing and modeling complex and long-range curvilinear geological features, which hinder the accurate capture of geological connections, led to the introduction of MPS algorithm. In the realm of advancing MPS implementations, it becomes evident that dedicated efforts are directed toward enhancing several pivotal facets. These include enhancing physical realism (spatial continuity), optimizing data conditioning, and boosting computational efficiency, particularly within large-scale systems [2]. Striking the appropriate equilibrium among these elements unveils new opportunities for addressing intricate problems.

In this research, we propose a new MPS implementation designed to rapidly simulate grids with millions of cells. Beyond its computational efficiency, the proposed methodology prioritizes the maintenance of spatial continuity and adherence to data conditioning principles to a considerable degree. To attain this objective, we incorporate CCF into the DWT to address some of the concerns mentioned earlier. Within our proposed method, the measurement of similarity between the TI patterns and data events takes place within the space of wavelet approximation coefficients. It is worth noting that each of these tools has been independently employed in geostatistical simulation [3-5]. However, in the current study we introduce a novel hybrid approach that combines these tools to enhance MPS simulation.

The structure of this work is as follows: In Section 2, we introduce earlier works related to the subject. In Section 3, we provide some background necessary for understanding our proposal. In Section 4, we provide a detailed explanation of the proposed method. Moving on to Section 5, we perform examinations, benchmarking, and in-depth analyses of the proposed method using various test cases. Finally, in Section 6, we engage in a discussion and draw our conclusions.

## 2. Related work

After the first non-iterative MPS algorithm [6], ENESIM, a more efficient extension of the original algorithm based on Single Normal Equation Simulation (SNESIM) was created [7]. This concept involved the incorporation of search trees to retain all multiple-point statistics from the TI. To improve pattern connectivity in realizations and accelerate algorithmic performance, pattern-based algorithms have been developed. These algorithms draw inspiration from the computer graphics community [8]. Pattern-based MPS techniques provide an alternative approach to deriving conditional probabilities for random variables at different points. Here, a 'pattern' refers to a set of values defined by a specific template. For example, it could be a 20 by 20 square in a 2D context over the TI grid. This method inherently examines the probabilities of entire multiple-point patterns, conditioned on the specific data events associated with those points in the TI. At each node location, the technique identifies the data event using a template and systematically explores all patterns in the TI to find the most similar one. Subsequently, the identified pattern is applied to the Simulation Grid (SG), and this iterative process continues until all nodes on the SG are visited. The similarity search for the data event and patterns takes into consideration the conditioning of neighboring data values, giving priority to a local pattern similarity criterion over local conditional distribution [9].

In the evolution of MPS algorithms, the transition from pixel-based to pattern-based simulation began with the introduction of the SIMPAT algorithm [9]. In a similar vein, a pattern-based algorithm, incorporating specific adjustments and amalgamating SIMPAT with texture synthesis concepts, has been developed [9, 10]. Their approach involves the gradual expansion of a randomly placed patch within the SG. To reduce computation time for similarity computation, they utilized DWT to compress extracted patterns from the TI. To further enhance simulation speed, some researchers proposed the concept of clustering patterns into a limited number of clusters [11-13]. Derived from Shannon [14], a direct sampling (DS) method has been developed for finding the desired pattern from the TI based on a pre-defined threshold, instead of searching through a pattern database [15]. To reduce the substantial computational workload associated with DS, the bunch pasting method was proposed, employing a strategy of pasting a group of nodes at each step [16].

Taking cues from the advancements in computer vision highlighted in the original study [17], CCF was used to calculate pattern similarity using a raster path for simulation [3]. Building on the concept of Image Quilting (IQ), minimal error boundary cuts between successive patches have been employed to enhance the spatial continuity of MPS simulation [18, 19].

### 2.1. Review of the MS-CCSIM (CCSIM)

In CCSIM algorithm, the need to construct pattern data from small data events was obviated. The algorithm introduced a novel similarity function to compare the generated pattern with the TI, employing CCF along a raster path. This function seamlessly integrated with effective strategies to uphold continuity and faithfully reproduce patterns. The primary objective was to discern a pattern or a set of patterns that aligned with a specific data event (OR) within a TI. This process hinged on the utilization of CCF, acting as a metric to gauge the similarity between two patterns. Additionally, there was a substantial reduction in CPU usage by a factor of 30–80 and a considerable decrease in the memory requirements for computations [3].

Subsequently, a fast extension of the CCSIM algorithm emerged, known as MS-CCSIM, designed specifically for the rapid simulation of categorical variables. This enhancement centered around the implementation of a multi-scale (MS) framework, diverging significantly from conventional multi-grid methods found in established methodologies. Within the proposed multi-scale framework, the original high-resolution TI underwent a transformation, resulting in a pyramid structure comprising progressively coarse-grid perspectives of the TI. This pyramid not only expedited the identification of patterns but also facilitated their seamless integration over a mutually shared overlapping area with previously simulated patterns. The multi-scale perspective not only showcased its effectiveness in data conditioning but also contributed to accelerating the conventional CCSIM algorithm [20].

Recently, there have been novel implementations leveraging cutting-edge computer vision approaches. These implementations excel in swiftly and effectively reproducing TI [21]. Regrettably, a considerable portion of these implementations either entirely neglect conditioning or suggest very restricted conditioning capabilities.

## 3. Background

Before delving into the core components of the proposed method, it is essential to revisit some fundamental concepts of DWT and CCF in the context of the present study.

### 3.1. Brief review of DWT

DWT is a signal processing multi-resolution decomposition technique [22]. It breaks a signal into sets in both the frequency and spatial domains through the convolution of low-pass and high-pass filters. When considering an image as a 2D signal, DWT extracts important directional and omnidirectional features in the form of low and high-frequency coefficients [23]. DWT decomposes a signal into shifted and dilated wavelet functions (mother wavelet) $\psi^B$ and a scaling function $\phi^{LL}$ forming an orthonormal basis in the square integrable Hilbert space $L^2$ ($\Re^2$) function [24]. For a given image, DWT can be computed at different decomposition levels denoted by $j$ ($j=1, 2, ..., J$). In each decomposition level, one omnidirectional approximation coefficient ($a$) and three directional detail coefficients ($H, V, D$) in the horizontal, vertical, and diagonal directions are provided. It is important to note that each of the four sub-bands has $\frac{1}{4^j}$ dimension of the original image, i.e., the number of dimensions depends on the decomposition level. If $TI(x,y)$ is considered as a 2D image with size $N \times N$ the DWT of TI at level $j$ is formulated by [24, 25]:

$$TI(x,y) = \sum_{k,i=0}^{Nj-1} a_{J,k,i}^k \phi_{J,k,i}^{LL}(x,y) + \sum_{B \in D} \sum_{j=1}^{J} \sum_{k,i=0}^{Nj-1} Z_{j,k,i}^B \psi_{j,k,i}^B(x,y),$$

where $B = \{H, V, D\}$, $\phi_{J,k,i}^{LL} \equiv 2^{\frac{-j}{2}} \phi(2^{-j}x - k, 2^{-j}y - i)$, $\psi_{j,k,i}^B \equiv 2^{\frac{-j}{2}} \psi^B(2^{-j}x - k, 2^{-j}y - i)$, $N_j = N/2^j$

The terms $a_{J,k,i} = \iint TI(x,y) \phi_{j,k,i} dxdy$ and $Z_{j,k,i}^B = \iint TI(x,y) \psi_{j,k,i}^B dxdy$ are approximation coefficients and detail coefficients at level $j$, respectively.

The main steps to compute 2D DWT on a discrete signal are:

- Convolution of image rows using low- and high-pass filters;
- Down-sampling on filtered image columns;
- Convolution of resulting image columns using same low- and high-pass filters;
- down-sampling on the resulting image rows.

Consequently, after these analyzing steps, four DWT sub-bands are produced at each level of decomposition. Fig. 1 illustrates a one level two-dimension DWT and its corresponding sub-bands. DWT approximation coefficients are capable of capturing the low-frequency variations within an image. Essentially, these coefficients encapsulate the most significant global variability present in the image. Furthermore, the inclusion of bijective filters, known as Inverse-DWT, enables a smooth transition from the wavelet feature space back to the original space [22]. To clarify this topic Fig. 2 illustrates a TI alongside its corresponding DWT coefficients at two different decomposition levels.

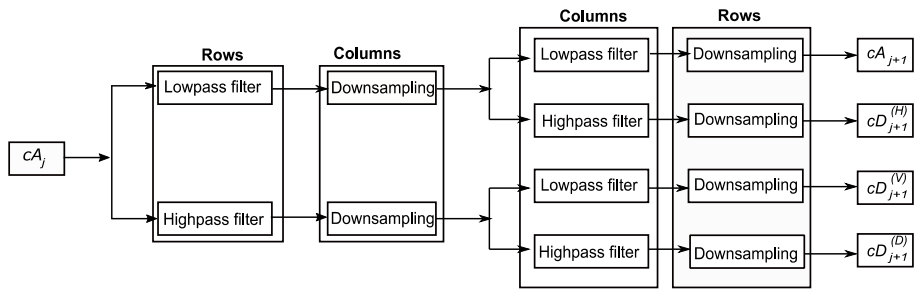

**Fig. 1**. One level 2D DWT on approximation coefficients ($cA_j$) obtained from level $j$ and four sub-bands at level $j+1$: approximation ($cA_{j+1}$), horizontal ($cD_{j+1}^H$), vertical ($cD_{j+1}^V$) and directional coefficients ($cD_{j+1}^D$) [22]

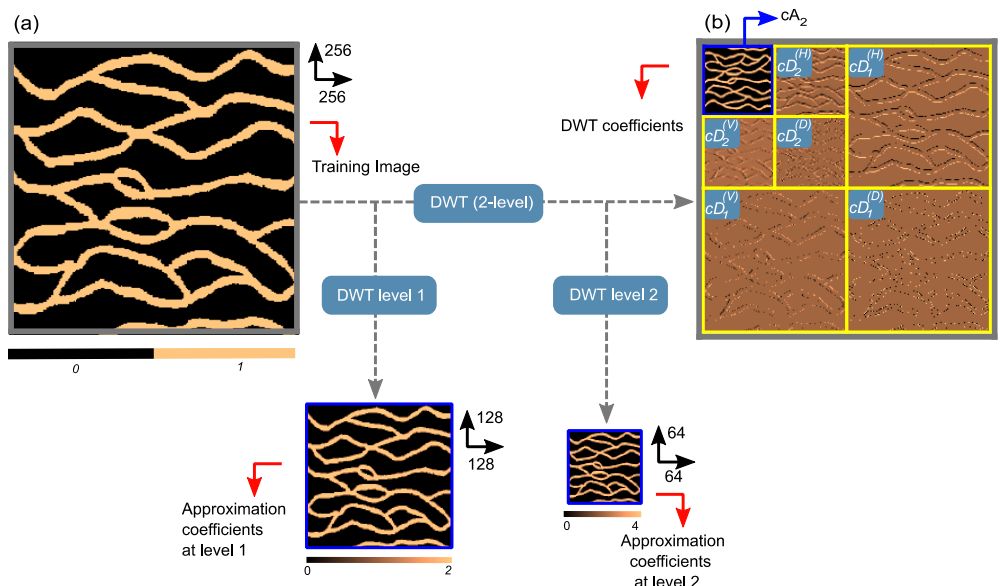

**Fig. 2**. Two levels of DWT: (a) the original image (b) approximation ($cA_2$) and detail coefficients ($cD^{(V,H,D)}$) at two different levels.

### 3.2. Combination of CCF and DWT

The CCF serves as a fundamental technique in computer vision for pattern matching. It aids in identifying similarities or matches between a template pattern and an image. The process involves multiplying the pixel values of the template and the image and then summing up the result. Peaks in the matching score map correspond to locations where the template is most similar to the image region. However, when dealing with large images or templates, direct computation of the correlation at every possible position can be time-consuming. This is because the time complexity of CCF is directly proportional to the product of the image size and the template size, leading to significant computational overhead when working with large inputs.

To tackle the aforementioned challenges, we present an innovative approach that merges the CCF with the DWT to pinpoint the most suitable pattern within TI. Following a single-level decomposition using the *Haar* mother wavelet, the approximation subband retains only a quarter of the total pixels from the original image. However, it's crucial to emphasize that the information within the approximation coefficients is not lost; instead, it is efficiently stored within the detail coefficients.

The concept behind the proposed method is to compute the CCF at a coarser level, utilizing the approximation coefficients from both the TI and the data event. This approach helps alleviate computational costs when compared to calculating the CCF at the original image resolution. The key advantage of this approach lies in the ability of approximation coefficients to capture low-frequency or global features that exist in both the TI and shared region in a pre-defined template called Overlapping Region (OR). Once a potentially matched pattern is identified within the approximation coefficients of the TI, it can be restored to its original form through a reverse transformation using the Inverse-DWT. A more detailed explanation of the proposed method will be provided in the subsequent sections.

## 4. The proposed method: CCWSIM

Within this section, we explore key concepts essential for gaining a comprehensive understanding of the proposed method. We begin by addressing the SG, raster path, OR. Following that, we delve into the details of conducting DWT for the compression of OR and TI. Subsequently, we elaborate on how the proposed method measures the similarity between patterns, identifies, and reconstructs desired patterns from approximation coefficients of TI. Fig. 3 illustrates the schematic workflow of the proposed method. In the following sections, we provide detailed insights into each step of this workflow.

### 4.1. Simulation grid and overlapping region

The first step involves establishing a lattice on the SG and a pre-defined raster path, which is determined by factors like dimensions of the SG, template size, and OR. The proposed method employs the concept of OR resembling the well-established image-quilting (IQ) technique used in texture synthesis [18]. To superimpose patterns along with the raster path, a small portion of the previously simulated pattern is defined as a data event (OR). The shape and size of OR change depending on the simulation's location, appearing as either an L-shaped or rectangular (vertical or horizontal). Three distinct scenarios for extracting OR unfold based on the location of the grids that are being simulated. Three potential scenarios in the proposed method are shown in Fig. 3a-c with blue regions. The simulation commences from the origin of the SG, progressing in a consistent trajectory either horizontally or vertically (see Fig. 3a). During the geostatistical simulation, best-matched patterns with OR (detailed in the section 4.3 and 4.4) are positioned along the raster path within the SG. This approach to navigating the SG pixels is inspired by the methodology employed in MPS algorithms [26].

### 4.2. Computing DWT

Once OR has been extracted from the SG, the subsequent task involves assessing the similarity between OR and the patterns within the TI. The computation of CCF takes place within the space of wavelet approximation coefficients. Therefore, it becomes essential to decompose both the TI and OR to a specified level of decomposition. Fig. 4 illustrates the process of conducting a 2-level DWT using the *Haar* mother wavelet on both the TI and OR, encompassing their respective approximation coefficients. The *Haar* mother wavelet basis function stands out as a particularly suitable choice for binary images. The *Haar* mother wavelet, along with its corresponding scaling function, has been defined as described by [22].

$$\psi^B(t) = \begin{cases} 1 & 0 \leq t \leq 1/2 \\ -1 & 1/2 \leq t \leq 1 \\ 0 & otherwise \end{cases} \quad , \quad \phi^{LL}(t) = \begin{cases} 1 & 0 \leq t \leq 1 \\ 0 & otherwise \end{cases}$$

### 4.3. Computing similarity using CCW

For a given TI and OR, we introduce a new criterion called CCW that measures the similarity between the TI patterns and OR patterns through approximation coefficients, which is formulated for pixel locations *(s,t)* as follows:

$$CCW(x,y) = \sum_{t=0}^{N_j-1} \sum_{s=0}^{M_j-1} cA_j^{TI}(x+s, y+t) \cdot cA_j^{OR}(s,t)$$

where $N_j = \frac{N}{2^j}, M_j = \frac{M}{2^j}$, $cA_j^{TI}$ and $cA_j^{OL}$ are the approximation coefficients of *TI* and *OR* at level *j*, respectively.

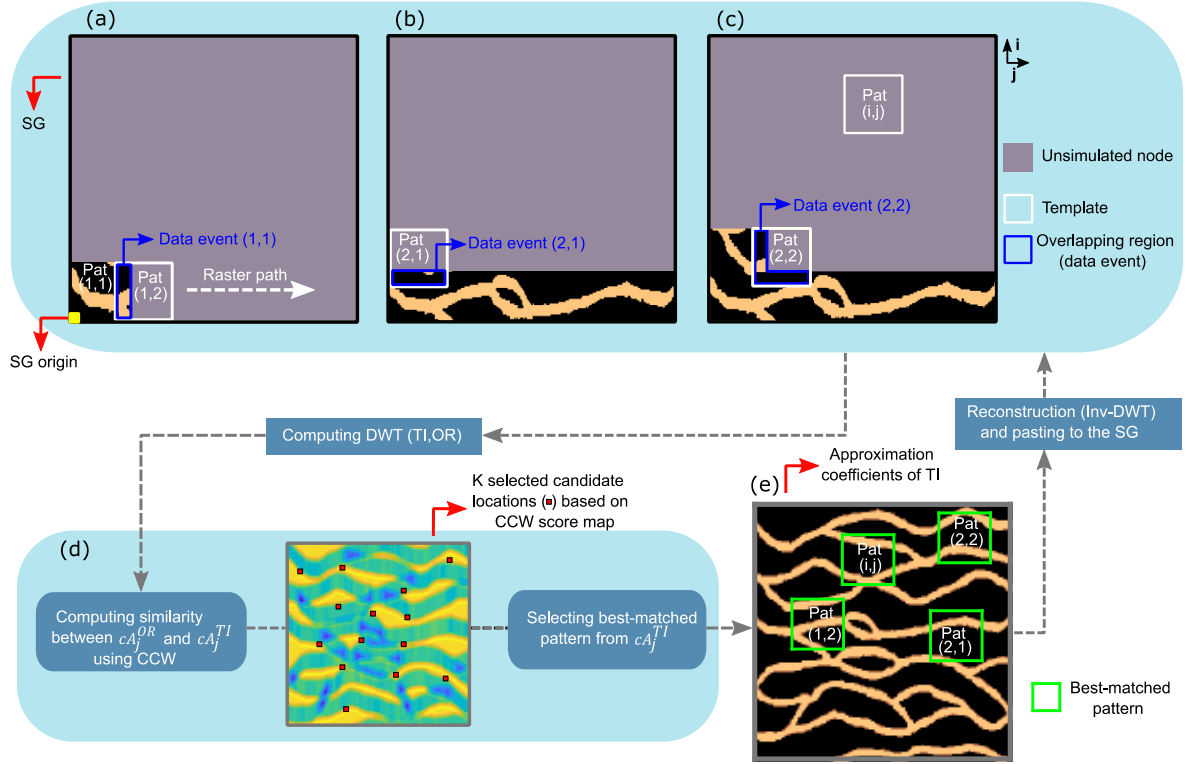

**Fig. 3**. A schematic illustration of the proposed method workflow. (a-c) The SG with three distinct OR scenarios from previously simulated patterns, (d) computing wavelet transform between TI and OR and (e) selecting best-matched pattern (For interpretation of the references to color in this figure legend, please refer to the online version of this article)

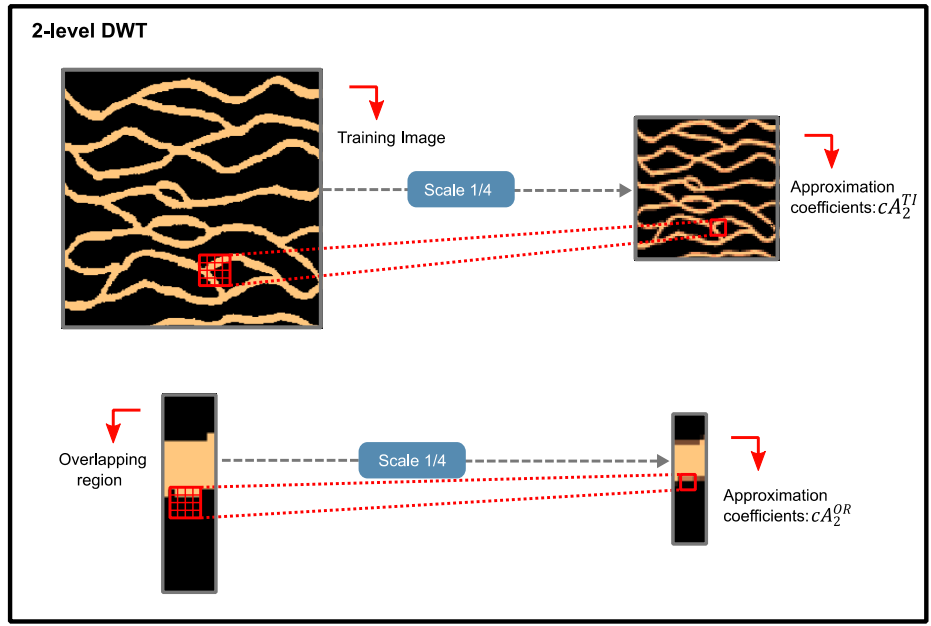

**Fig. 4**. Conducting 2-level DWT on both the TI and OR, along with their respective wavelet approximation coefficients.

The size of TI after undergoing $j$ level(s) of decomposition is denoted as $M_j \times N_j$. Consequently, the size of the original pattern can be significantly reduced depending on the wavelet decomposition level. For instance, consider a 2D TI with dimensions of 2000 × 2000 (4,000,000 pixels). When subjected to the 3 levels ($j=3$) of wavelet decomposition, the vector's length shrinks from 4,000,000 pixels to 62,500 pixels (250 × 250), representing a substantial reduction. This computational advantage also extends to OR.

The key feature introduced by the proposed method is the integration of CCF with DWT to measure the similarity between patterns. In contrast to conventional interpolation techniques such as bicubic interpolation, the proposed method utilizes DWT for feature extraction and compression of the pattern. After computing the approximation coefficients of OR (data event) and TI ($cA_j^{TI}$) at a specific DWT level, their similarity is measured through CCW. During this process, CCW score map is examined to identify locations of patterns with the highest similarity to the approximation coefficients of OR ($cA_j^{OR}$) using the CCW score (see Fig. 3d).

### 4.4. Identifying and reconstruction a best-matched pattern

After calculating the CCW score map, the pattern selection process differs depending on the simulation type, whether it is unconditional or conditional. In unconditional simulation, the procedure entails sorting the CCW scores, selecting a specific number of top-ranked candidates' locations, and then performing a random draw to determine the location of the best-matched pattern. The pseudocode provided in Algorithm 1 summarizes the unconditional simulation through the proposed method. In line 2, DWT of the TI is computed based on a user-defined level ($j$). To introduce variability between realizations, variations in the origin of the raster path are implemented (line 3). At each location $l = (x, y)$, if no simulated pattern exists in the SG, a pattern is randomly selected from the TI (lines 5 and 6). Subsequently, OR is extracted from the previously simulated pattern, and its DWT at level $j$ is computed (lines 8 and 9). Moving to line 10, CCW score between the approximation coefficients of the TI and OR are computed. Based on the score map, K top-ranked candidates are selected, and one candidate pattern is randomly drawn (lines 11 and 12). Finally, the Inverse-DWT of the selected pattern, using detail coefficients, is computed, and the reconstructed (original) pattern is pasted into the SG (lines 13 and 14). Fig. 5 provides a visual illustration of the identification and reconstruction of best-matched patterns in the proposed approach.

Algorithm 1

| Pseudocode for unconditional simulation using the proposed method |
|---|
| **Require**: **TI**= Training Image |
| **Require**: **T**$_{size}$ = Template size |
| **Require**: **OR**$_{size}$ = Overlapping Region size |
| **Require**: **J** = Wavelet decomposition level |
| **Require**: **K**= Maximum candidate patterns |
| **Require**: **n**$_r$= Number of realizations |
| 1: **for** r = 1 to **n**$_{real}$ **do** |
| 2:     $cA_J^{TI}$ ← compute DWT of **TI** at level $J$ |
| 3:     simulation_path ← Randomize(origin, direction) |
| 4:     **for** each location $l$ in the SG **do** |
| 5:         **if** there is no simulated patch in the SG **then** |
| 6:           select randomly a patch from **TI** |
| 7:         **else** |
| 8:           **OR**$_l$ ← extract ovelapping region |
| 9:           $cA_J^{OR}$ ← compute DWT$_J$( **OR**$_l$) |
| 10:           **CCW**$_l$ ← compute CCW( $cA_J^{TI}$ , $cA_J^{OR}$) |
| 11:           **cand_pattren** ← find **K** best candidate patterns |
| 12:           **sel_pattern** ← drawn randomly (**cand_pattren**) |
| 13:           **rec_pattern** ← Inv − DWT(**sel_pattern**) |
| 14:           SG($l$) ← paste(**rec_pattern**) |
| 15:         **end if** |
| 16:     **end for** |
| 17: **end for** |

For conditional simulation, a methodology reminiscent of that introduced by [20] is adapted. In this context, a virtual template encompassing the Template (T) is considered and partitioned into two distinct segments: a causal region that governs the conditional location within T, and a non-causal region that incorporates conditional data originating from a pioneering region within the virtual template. Therefore, after the similarities (CCW) are sorted, a sequential search strategy is employed to identify the location of a best-matched pattern, while simultaneously ensuring the preservation of conditional data within both the causal and non-causal regions. Further insights into the data conditioning process can be found in the original paper [20].

## 5. Experiments

All simulations in this research were performed using MATLAB on a laptop equipped with an Intel Core i7 processor running at 2.6 GHz and 16 GB of RAM.

### 5.1. Unconditional simulation

In the initial experiment, we utilized the proposed method to create unconditional realizations using the Channel TI. Fig. 6 displays the Channel TI, simulation input parameters, and various unconditional realizations generated through the proposed method. The realizations appear to be satisfactory in terms of preserving the pattern and overall continuity.

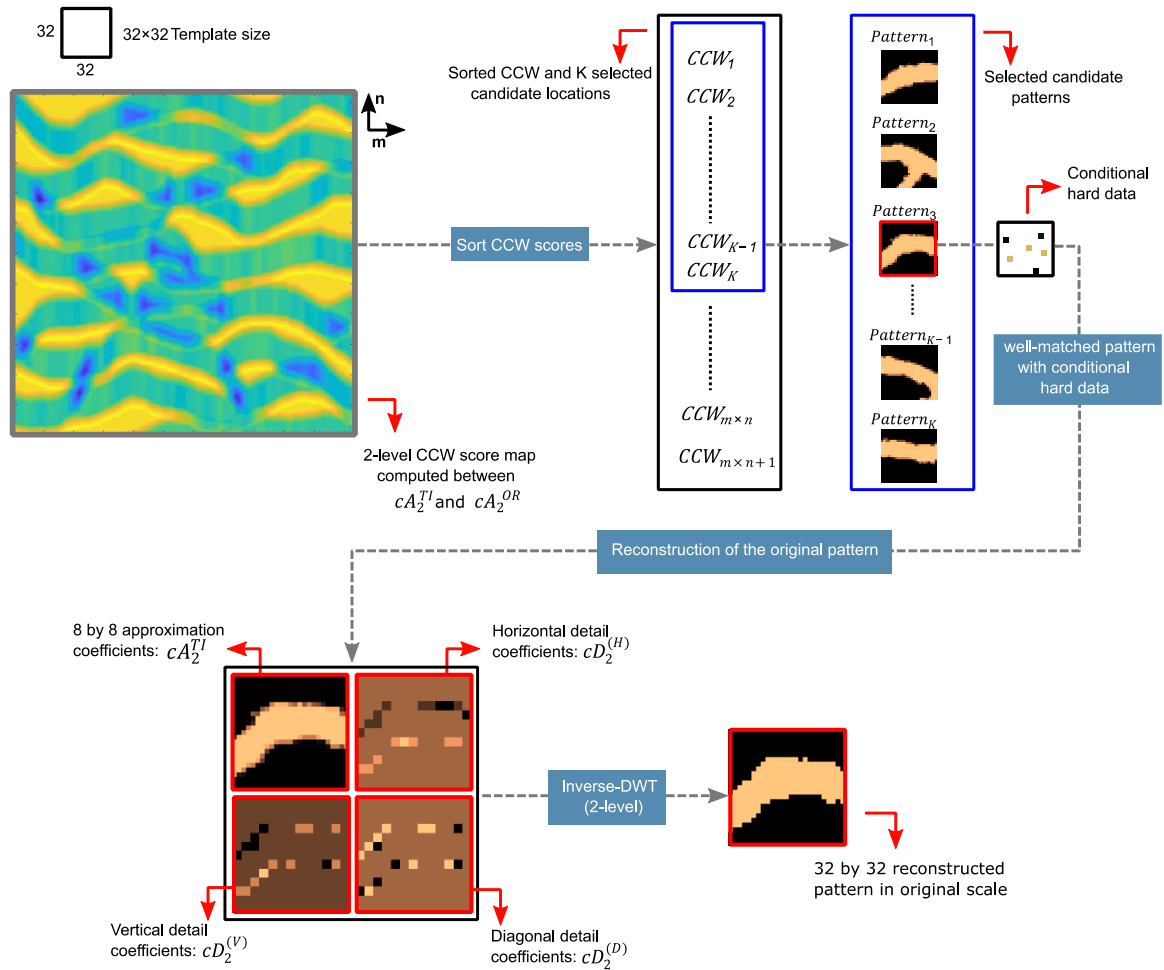

**Fig. 5**. Schematic illustration of the identification and reconstruction of best-matched patterns using the CCW score map in the proposed method.

## 5.2. Conditional simulation

Data conditioning, particularly hard data conditioning in geostatistical simulations, is a crucial aspect that involves integrating and honoring real-world data collected in the field. In the following sections, we assess the performance of the proposed method in various conditional simulations. To ensure consistent conditioning data with matching histogram and variogram patterns for all conditional simulations in this study, the process starts by generating an initial unconditional realization. Then, random samples are selected from this initial realization as hard data.

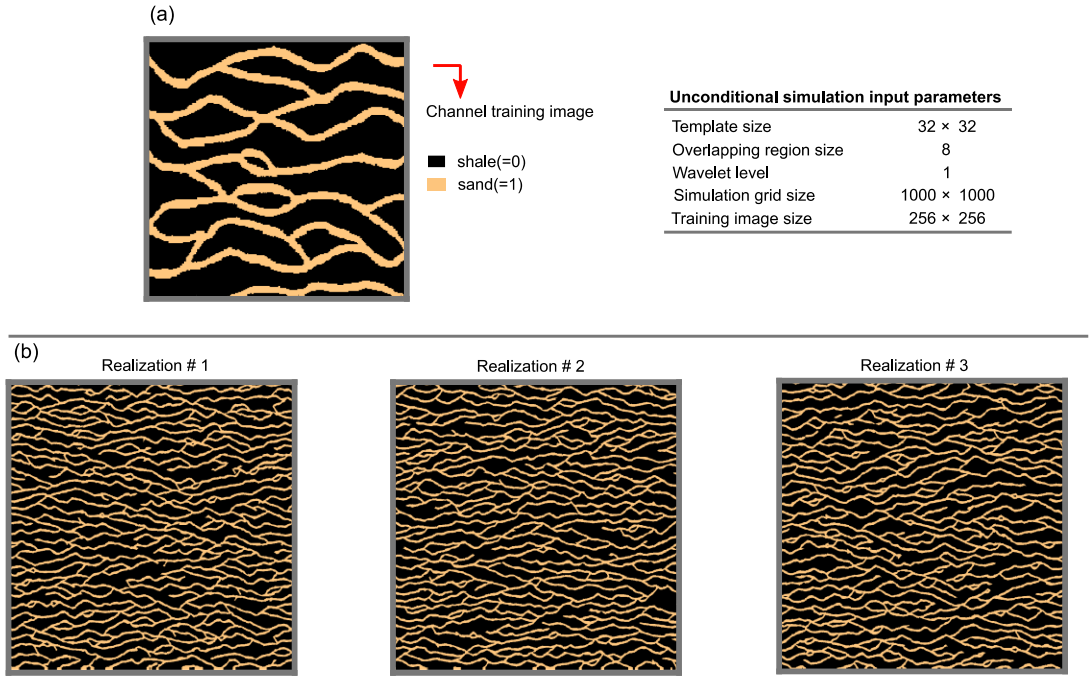

**Fig. 6**. (a) The Channel TI [7] and (b) three unconditional realizations obtained using the proposed method (average CPU time: 6.52 seconds).

A set of 100 realizations was generated using the TI displayed in Fig. 6a. Three individual realizations, simulation input parameters and ensemble average map of 100 realizations are depicted in Fig. 7a, b, respectively. The conditional realizations not only faithfully honor the provided conditioning data but also strive to preserve the connectivity of the channels. Additionally, the ensemble average map in the vicinity of the conditional locations highlights the structure of the channels. On average, the proposed method has attained convergence in locations near the conditional hard data.

The performance of the proposed method was assessed using a different TI that incorporates facies with a higher degree of complexity, derived from a satellite image of the Ganges delta in Bangladesh. Fig. 8 displays the TI, various conditional realizations generated using the proposed method, and the locations of the conditional hard data. The outcomes indicate that the proposed method is well-suited for reproducing complex pattern structures, preserving continuity, and adhering to hard data conditioning.

## 5.3. Validation of the results

In this section, we validated the simulation results of the proposed method using two metrics: the variogram and the probability of connection. The graphical representation of these metrics is shown in Fig. 9a, b, encompassing the variograms and the probability of connection associated with both the Channel TI in Fig. 6a and the set of 100 realizations generated through the proposed method. Variograms were computed along two cardinal axes: the North-South and East-West directions. Because of the dominant continuity of channels in the East-West orientation, the evaluation of the probability of connection was limited to this specific direction. It appears that the proposed approach allows reaching an acceptable level of consistency in reproducing the variograms and the probability of connections with the TI, particularly within the specified direction.

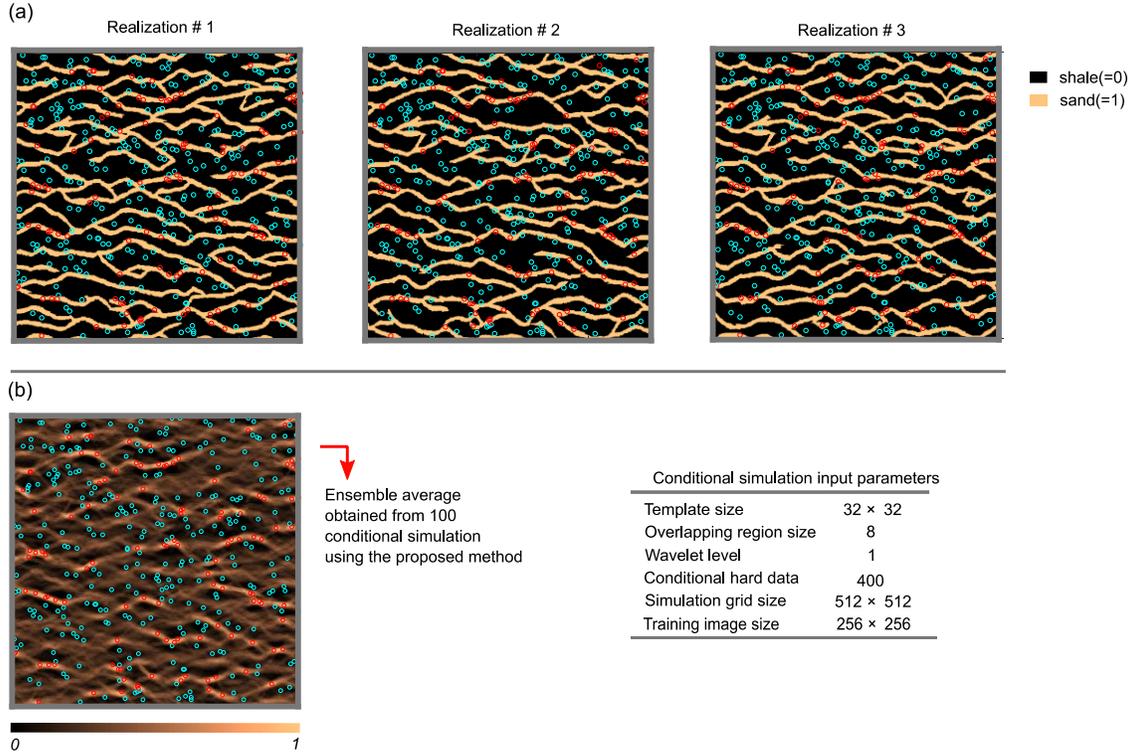

**Fig. 7**. Conditional simulations using the proposed method: (a) three distinct conditional simulations and (b) the ensemble average map obtained from 100 conditional realizations using the proposed method (average CPU time: 2.42 seconds). Conditional hard data are indicated by blue and red circles.

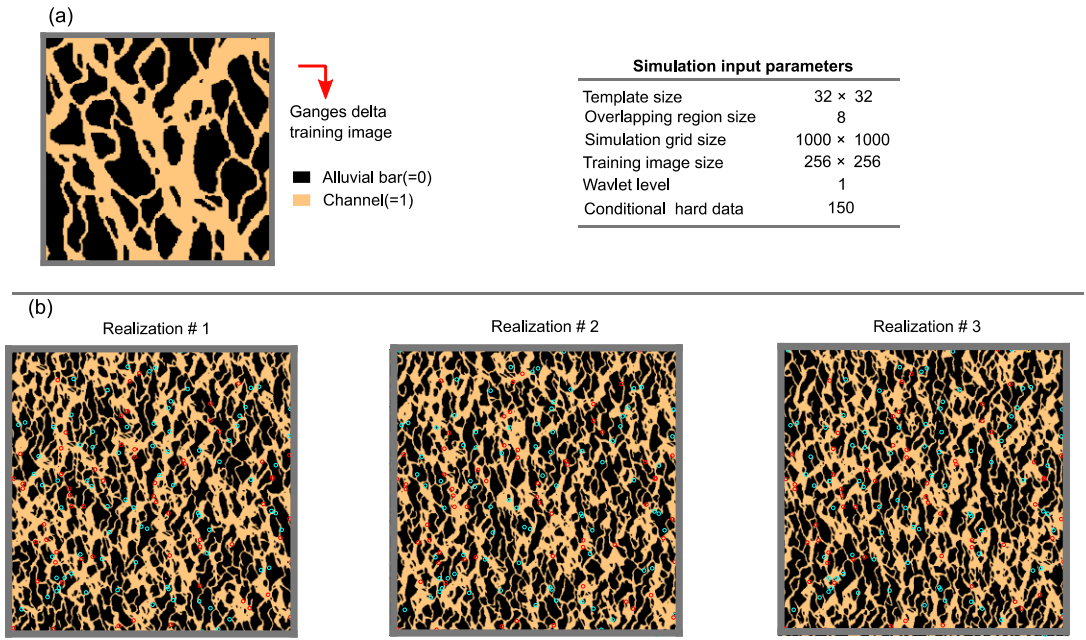

**Fig. 8**. Conditional realizations of the Gange delta TI obtained using the proposed method (average CPU time: 7.7 seconds). The blue and red circles mark the positions of the conditional data. (For interpretation of the references to color in this figure, the reader is referred to the web version of this article)

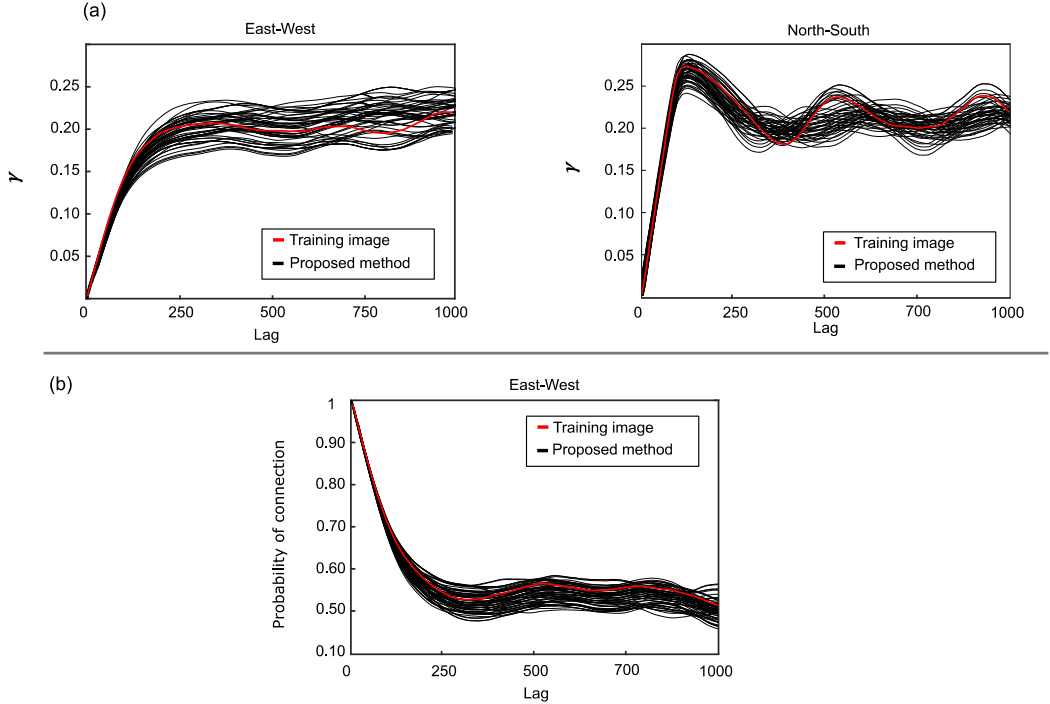

**Fig. 9**. (a) The variograms and (b) the probability of connection for the Channel TI and 100 realizations obtained from the proposed method.

### 5.4. Benchmarking

In this section, we evaluated the performance of the method by benchmarking it with MS-CCSIM algorithm, which is available at https://github.com/SCRFpublic/MS_CCSIM [20]. MS-CCSIM was specifically developed to enhance computational efficiency for large-scale categorical systems. Both MS-CCSIM and the proposed method share the common objective of expediting the simulation process by searching for desired patterns within a coarse-grid TI. The first comparative study involves generating conditional simulations using Fig. 6a at various multi-scale and wavelet levels. Additionally, different sets of conditional hard data were introduced into the conditional simulation to assess their impact on the simulation results. Comparative results for the proposed method and MS-CCSIM, obtained from different multi-scale levels, wavelet levels, and conditional hard data, are presented in Fig. 10. Upon visual inspection, it appears that the proposed method generates patterns with spatial continuity that align more with the TI.

In the next comparative experiment, conditional realizations were generated using a categorical version of the Stonewall TI (Fig. 11a). For this experiment, 300 conditional hard data points were extracted from a reference image. Conditional simulation, input parameters and different realizations generated through MS-CCSIM and the proposed method are shown in Fig. 11. It appears that the proposed method produces patterns with fewer discontinuities when compared to MS-CCSIM.

### 5.5. Quantitative evaluation

By using Analysis of Distance (https://github.com/SCRFpublic/ANODI) introduced by [27], one can quantitatively evaluate the performance of MPS implementations. This metric is constructed based on two key components: between-realization variability and within-realization variability. Between-realization variability measures the spatial uncertainty within a set of realizations, while within-realization variability assesses the algorithm's ability to reproduce the patterns present in the TI. These measures are presented as a ratio described in the following equation:

$$r_{(A,B)} = \sum_{p=1}^{P} w_p \frac{d_{(A,B)}^{between}}{d_{(A,B)}^{within}}$$

where $w$ is the corresponding weights at the level of resolution $p$.

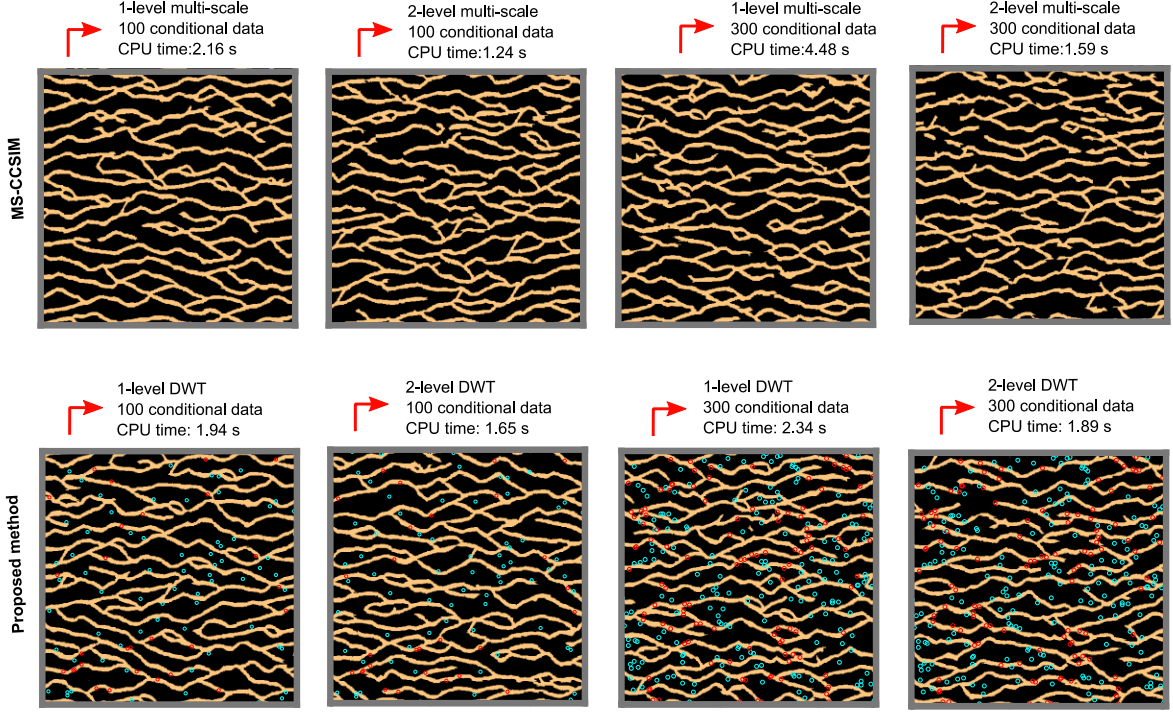

**Fig. 10**. Conditional simulations obtained through the proposed method and MS-CCSIM algorithm using different multi-scales, DWT levels and conditional hard data. The positions of the conditional hard data in the proposed method are indicated by circles in both blue and red colors.

If algorithm A performs better in terms of spatial uncertainty, then $d_{(A,B)}^{between}$ will be greater than 1. Conversely, if it excels in pattern reproduction, $d_{(A,B)}^{within}$ will be less than 1. Therefore, $r_{(A,B)}$ greater than 1 indicates that algorithm A outperforms algorithm B.

In the present study, between-realization variability is computed by averaging Jensen-Shannon divergence values pairwise among the 50 realizations across three resolution levels and 8 × 8 window size. In addition, within-realization variability is determined by averaging Jensen-Shannon divergence values between the 50 realizations and the TI across these same three resolution levels. Quantitative comparisons are conducted between MS-CCSIM and the proposed method using different TIs are shown in Fig. 6a and Fig. 11a. The results, which encompass both within and between variability, for three resolution level are summarized in Table 1. In most cases, $d_{(A,B)}^{between}$ falls below 1, indicating that MS-CCSIM exhibits greater variability between realizations (spatial uncertainty). Conversely, within-realization variability ratios are lower than 1, signifying that the proposed method more faithfully reproduces the patterns of the TI. The $r_{(A,B)}$ (where A and B represent the proposed method and MS-CCSIM, respectively) surpass 1 (1.01 and 1.19) for the Channel and Stonewall TIs, respectively. This suggests that the proposed method outperforms MS-CCSIM, especially in the case of the Stonewall TI.

In addition, for visual inspection Fig. 12a, b displays the Multidimensional Scaling (MDS) of 50 realizations for Fig. 6a and Fig. 11a, respectively . In each plot, the black dot denotes the TI, while the blue and green points represent realizations generated with MS-CCSIM and the proposed method, respectively. The axes were omitted to emphasize the relative distances between points.

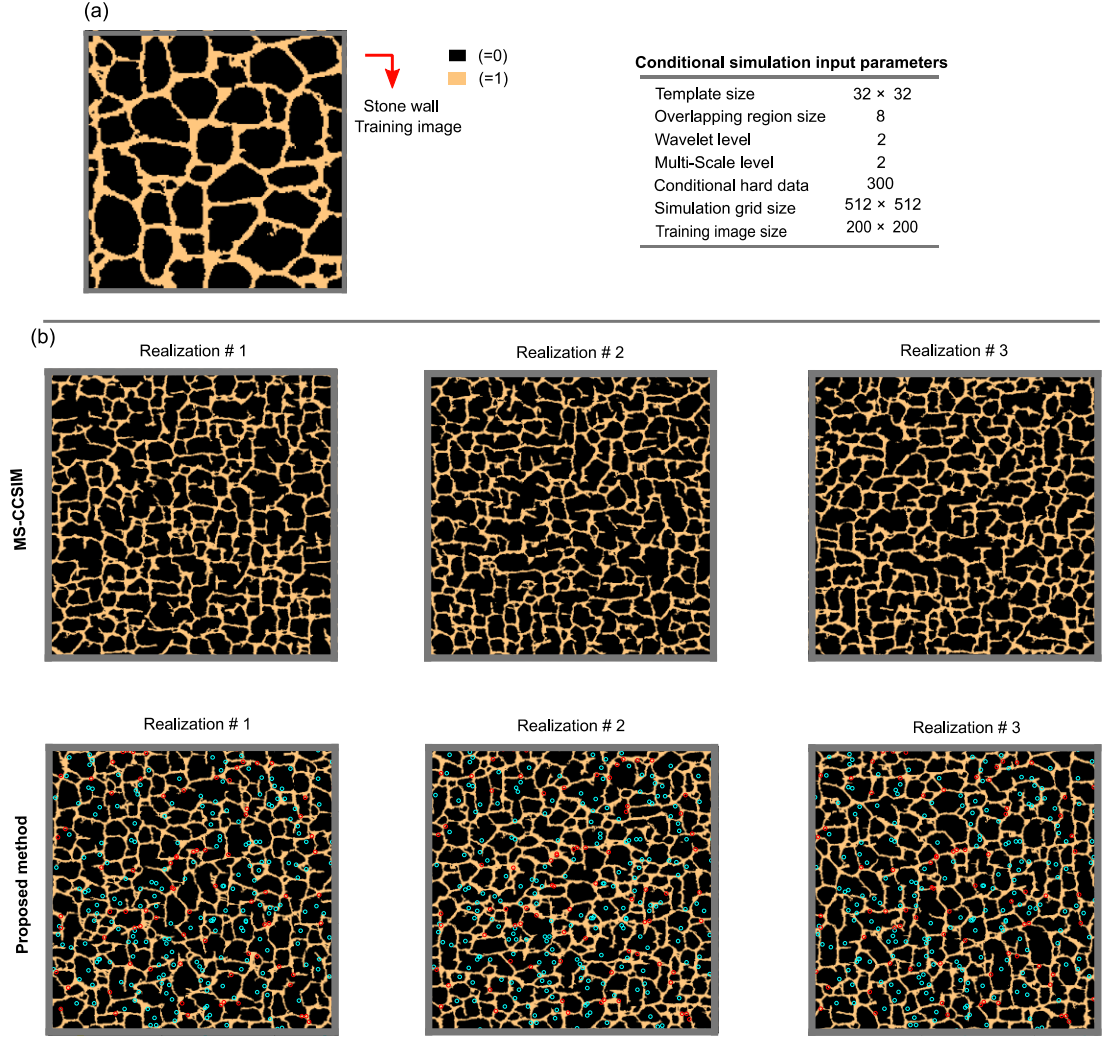

**Fig. 11**. Conditional simulations obtained from the proposed method and MS-CCSIM algorithm. In the proposed approach, circles of both blue and red colors denote the locations of the conditional hard data. (For interpretation of the references to color in this figure legend, please refer to the online version of this article)

### 5.6. Quantifying computational gain

To evaluate the computational efficiency of the proposed approach, we conducted a comprehensive study. In the first experiment in this section, the average CPU times for 20 conditional simulations of the TI in Fig. 6a, considering three different DWT levels and five SG sizes were recorded, as depicted in Fig. 13. As the DWT level increased from 1 to 2,

**Table 1** Between and within-realization variability ratio for three resolution levels obtained from the proposed method and MS-CCSIM.

| TI | $d_{(A,B)}^{between}$ | | | $d_{(A,B)}^{within}$ | | | $r_{(A,B)}$ |
|---|---|---|---|---|---|---|---|
| | Level 1 | Level 2 | Level 3 | Level 1 | Level 2 | Level 3 | |
| Fig. 11a | 0.95 | 0.92 | 1.02 | 0.79 | 0.79 | 0.95 | 1.19 |
| Fig. 6a | 0.79 | 0.78 | 0.88 | 0.78 | 0.77 | 0.99 | 1.01 |

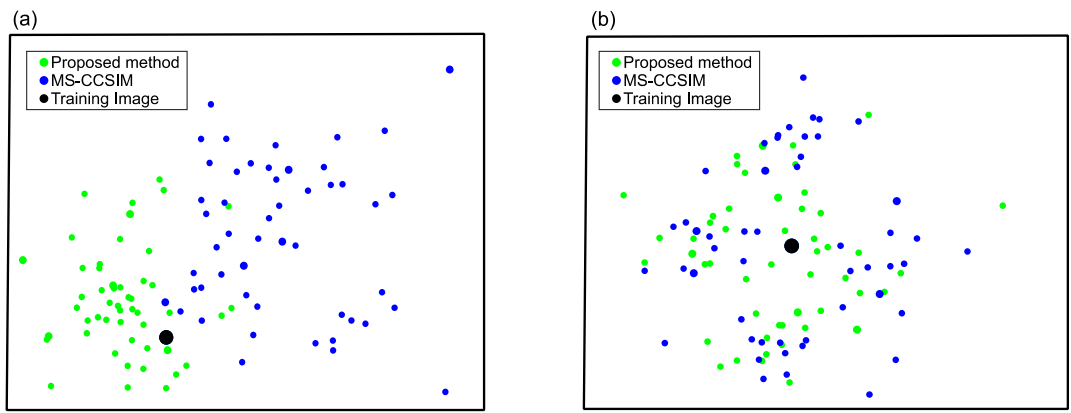

**Fig. 12.** MDS plots to illustrate the variability and pattern reproduction in both MS-CCSIM and the proposed method realizations, (a) Channel (Fig. 6a) and (b) Stonewall (Fig. 11a). (For interpretation of the references to color in this figure, the reader is referred to the web version of this article)

the CPU times decreased on average by 54.8% across various SG sizes. This reduction in computational cost after each DWT level follows an exponential trend, owing to the exponential nature of dimension reduction. Therefore, the improvement in computational efficiency from the second DWT level to the third DWT level is notably lower (17.4%) than the efficiency gain observed when transitioning from the first DWT level to the second DWT level.

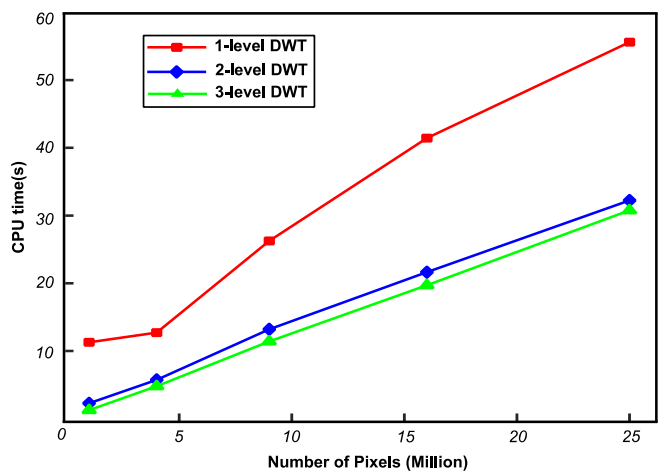

**Fig. 13.** Average CPU times for five different SG sizes using the TI in Fig. 6a at different DWT levels (1, 2 and 3).

In the next experiment, we extensively compared CPU times between the proposed method and MS-CCSIM (CCSIM), considering various TIs and related parameters such as multi-scale level, DWT level, and conditional data. For each scenario, we recorded average CPU time for simulating 20 conditional realizations using two sets of conditional data (1000 and 2000 sample points) for Channel (Fig. 6a), Stonewall (Fig. 11a), and Gang delta (Fig. 8a) TIs.

**Table 2** Comparison of average CPU time (s) for MS-CCSIM (CCSIM) and the proposed method using different TIs, muli-scale, DWT levels and conditional data (CD).

| Algorithm | Channel | | Stonewall | | Gange delta | |
|---|---|---|---|---|---|---|
| | 1000 CD | 2000 CD | 1000 HD | 2000 CD | 1000 HD | 2000 CD |
| MS-CCSIM (level 1) | 26.78 | 41.18 | 14.53 | 21.62 | 30.57 | 42.52 |
| MS-CCSIM (level 3) | 2.35 | 3.42 | 1.93 | 3.05 | 2.52 | 3.35 |
| Proposed (level 1) | 10.45 | 23.09 | 9.27 | 16.76 | 14.71 | 28.18 |
| Proposed (level 3) | 2.49 | 2.82 | 2.3 | 2.45 | 2.75 | 2.91 |

The realizations have a grid size of 512 × 512 pixels, generated with a template size of 32 × 32 and an OR size of 8. Table 2 presents the average CPU time in seconds for MS-CCSIM (CCSIM) and the proposed method.

### 5.7. Dense data conditioning

In essence, the proposed method involves selecting best-matched patterns from a coarse-scale TI within a compact space. In situations characterized by dense data, there may be concerns about the algorithm's ability to maintain pattern continuity. To address this concern, we conducted conditional simulations using 4,000 and 8,000 hard data points sampled from reference images to assess the effectiveness of the proposed method in the context of dense hard data. Fig. 14 and Fig. 15 illustrate the reference images and three distinct realizations generated by the proposed method, along with the TI in Fig. 6a and Fig. 11a. While some discontinuity artifacts may be observed in the realizations, it is evident that the proposed method demonstrates a tendency to preserve pattern connectivity even in the presence of dense hard data.

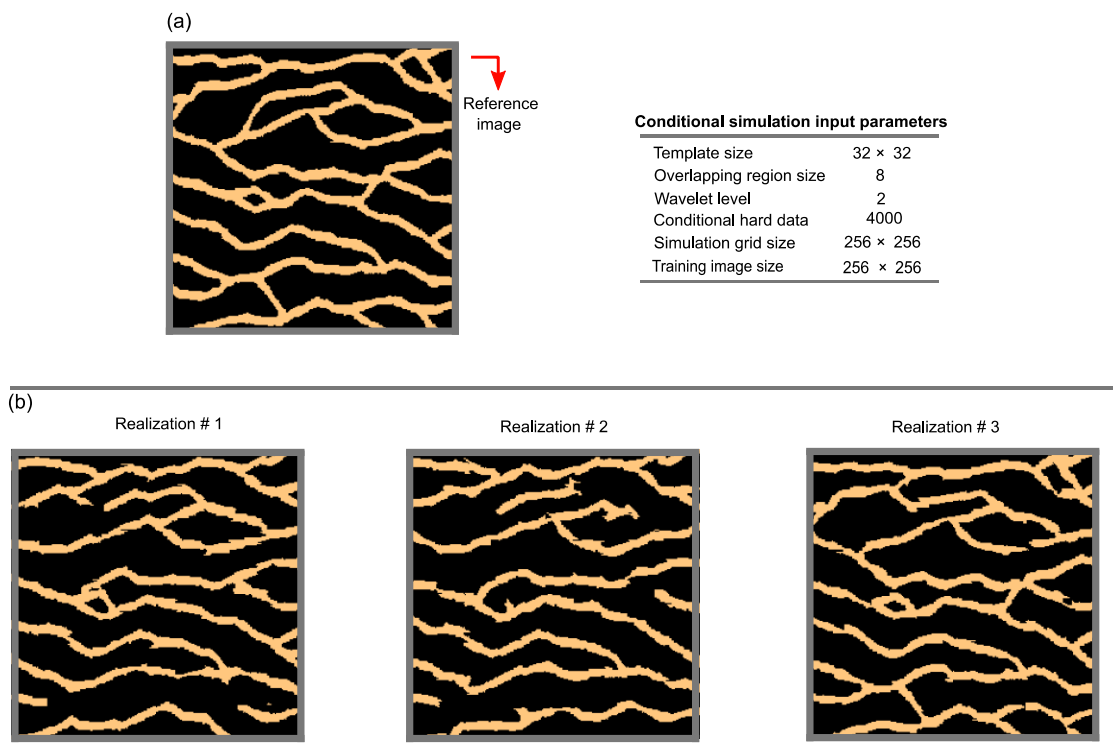

**Fig. 14.** (a) Reference image for Channel TI and (b) three conditional simulations generated using the proposed method (average CPU time: 1.06 seconds).

### 5.8. Muti-facies TI

In the next experiment, a multi-facies TI was chosen to validate the proposed method incorporating multivariable TI. 100 conditional simulations utilizing the TI showed in Fig. 16a were generated. The simulation parameters, the reproduction of facies proportion (%) in 100 realizations (solid lines) and TI (dotted lines), as well as three conditional realizations of the multi-facies TI obtained through the proposed method, are illustrated in Fig. 16b, c. The average CPU time for a 1000×1000 SG size was recorded at 3.50 seconds using 2-level DWT. Notably, the proposed method demonstrated the capability to achieve an acceptable quality, signifying its proficiency in accurately representing the characteristics of the multi-facies TI.

### 5.9. Nonstationary TI

In the last example, we evaluated the performance of the proposed method in the case of a nonstationary TI. A concern arises because the proposed method utilizes DWT approximation coefficients for computing the similarity between Bangladesh was selected. The nonstationary TI, ensemble average of 100 conditional simulations, and three different patterns, which poses challenges in simulating nonstationary TIs containing structures on several scales and orientations. To clarify this limitation, a categorical version of a satellite image depicting the Sundarbans region in realizations are shown in Fig. 17. It seems that the proposed method faced challenges in reproducing fine-scale structures on the TI; however, the realizations and ensemble average indicate a level of agreement with TI patterns. Nonetheless, it is important to note that the proposed method is an extended version of the original CCSIM, which was not specifically designed for simulating nonstationary TIs.

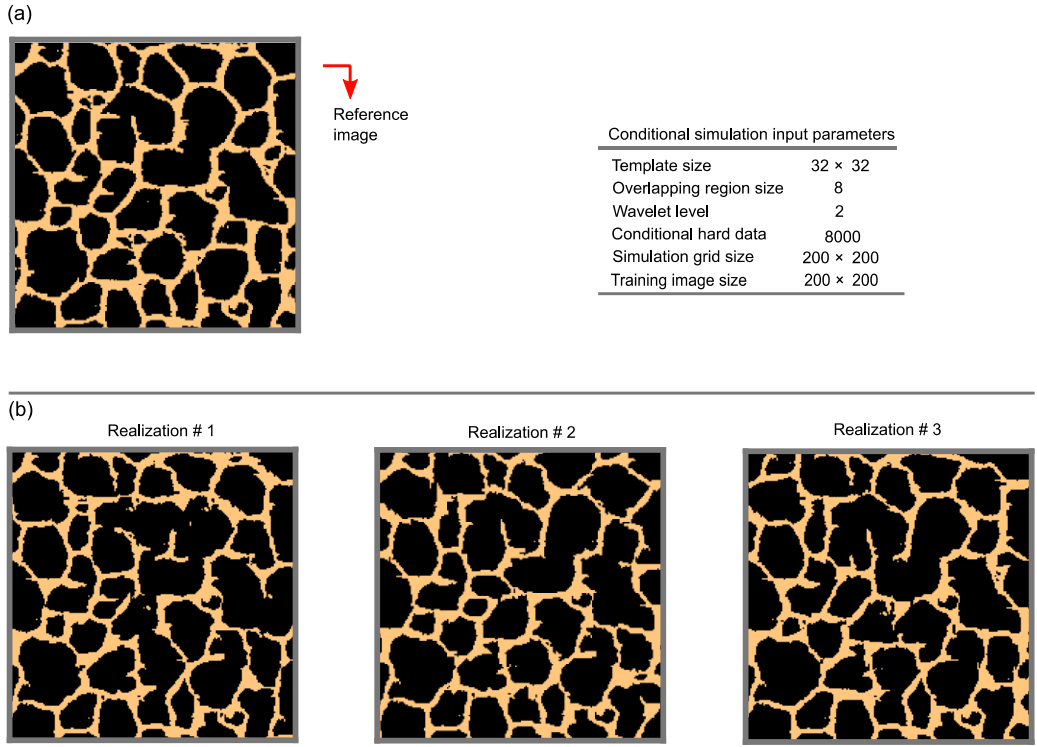

**Fig. 15**. (a) Reference image for the Stonewall TI and (b) three conditional simulations generated using the proposed

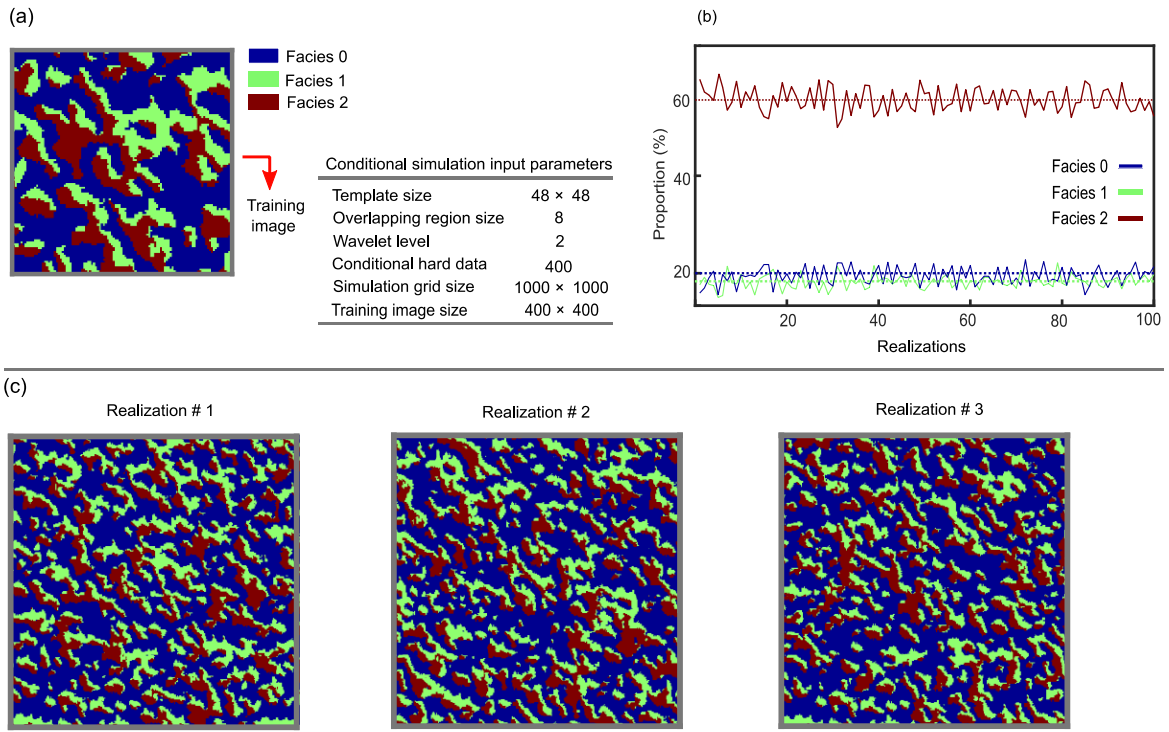

**Fig. 16**. (a) A Muti-facies TI, (b) The reproduction of facies proportion (%) in 100 realizations (solid lines) and TI (dotted lines) and (c) three conditional realizations of the muti-facies TI obtained using the proposed method.

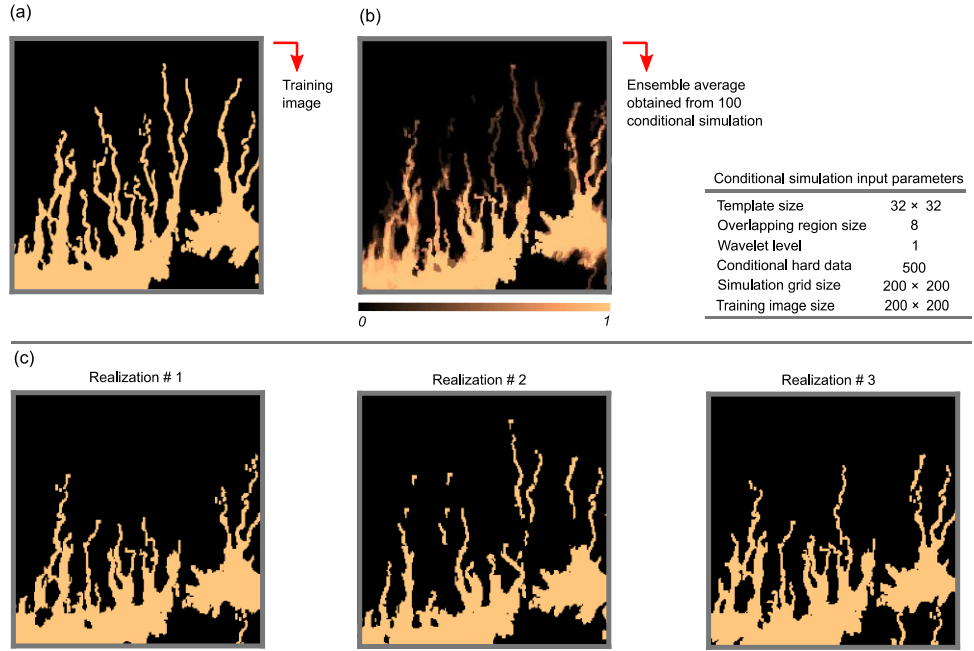

**Fig. 17** Conditional simulations using nonstationary TI (a) a categorical satellite image showing the Sundarbans, Bangladesh, (b) the ensemble average map obtained from 100 conditional realizations and (c) three distinct conditional simulations using the proposed method.

## 6. Discussion and conclusion

In this research, we introduced a wavelet-based implementation of the CCSIM algorithm for characterizing large-scale categorical systems. By computing the CCF within the DWT coefficient space, we achieved a notable reduction in CPU time, especially when transitioning from the first level of decomposition to the second.

While 75% of the details in the original pattern are discarded at each level of DWT decomposition, the evaluation results reveal a favorable balance between efficiency and accuracy in our proposed method. Quantitative comparisons indicate that the proposed method generates realizations on par with MS-CCSIM, with slight enhancements in reproducing TI patterns, at a lower or comparable computational cost.

An intriguing feature of approximation coefficients is their ability to retain potential matching patterns effectively after DWT decomposition, all while achieving significant dimension reduction. Essentially, what sets the proposed method apart is its departure from conventional interpolation techniques (like bi-cubic) for generating coarse-scale patterns. Instead, it utilizes DWT as an efficient tool for feature extraction.

Finally, from an industrial standpoint, it is noteworthy that no commercial implementation currently exists, given the research-oriented nature of this work. Nonetheless, one could endeavor to implement the algorithm in the C programming language, extending it into a three-dimensional (3D) context. This undertaking is feasible due to the utilization of wavelet transformation applicable in 3D domains. Such an implementation would assist practitioners in achieving a fast MPS implementation for characterizing large-scale geological domains with fewer artifacts and discontinuities.

# 7. Acknowledgments